# Ground States of the Hydrogen Molecule and Its Molecular Ion in the Presence of Magnetic Field Using the Variational Monte Carlo Method

**S. B. Doma[1)], M. Abu-Shady[2)], F. N. El-Gammal[2)] and A. A. Amer[2)]**

[1)] Mathematics Department, Faculty of Science, Alexandria University, Alexandria, Egypt
E-mail address: sbdoma@yahoo.com
[2)] Mathematics Department, Faculty of Science, Menofia University, Shebin El-Kom, Egypt

**Absract**

By Using the variational Monte Carlo (VMC) method, we calculate the $1s\sigma_g$ state energies, the dissociation energies and the binding energies of the hydrogen molecule and its molecular ion in the presence of an aligned magnetic field regime between $0\ a.u.$ and $10\ a.u.$. The present calculations are based on using two types of compact and accurate trial wave functions, which are put forward for consideration in calculating energies in the absence of magnetic field. The obtained results are compared with the most recent accurate values. We conclude that the applications of VMC method can be extended successfully to cover the case of molecules under the effect of the magnetic field.

## 1. Introduction

The influence of a magnetic field on the properties of molecules is of great interest. The simplest molecules, the hydrogen molecule $H_2$ and its molecular ion $H_2^+$, allow us to study their properties in strong magnetic fields with high accuracy. This molecule–field system has an essential importance not only in atomic and molecular physics, but also in astrophysics and semiconductor physics, solid state and plasma physics [1].

The behaviour of the $H_2^+$ molecular ion and the hydrogen molecule $H_2$ under strong magnetic field conditions has been studied by many authors. Most of them deal with the hydrogen molecular ion $H_2^+$ and little is known about the hydrogen molecule $H_2$. Research on the behavior of molecules in strong fields is more complicated than that of atoms because of the multi-centre characteristics of the molecules. The total energies and the equilibrium internuclear separations of $H_2^+$ molecular ion in the states $\sigma_g$, $\pi_u$, $\delta_g$, $\phi_u$, $\gamma_g$, $\eta_u$ in strong magnetic fields have been calculated using the adiabatic approximation and adiabatic variational approximation with an effective potential by Yong et al. [2]. Using the two-dimensional pseudospectral method, the ground and low-lying states of the $H_2^+$ molecular ion in a strong magnetic field are calculated in Ref [3]. The hydrogen molecular ion $H_2^+$ aligned with a magnetic field has been studied with the Lagrange-mesh method which allows obtaining highly accurate results under various field strengths and for various quantum numbers [4]. Turbiner et al. [5] studied the qualitative and quantitative consideration of the one-electron molecular systems such as $H_2^+$, $H_3^{2+}$ and $H_4^{3+}$ in the presence of a magnetic field. Using an accurate one center method with a technique that combining the spheroidal coordinate and B-spline, Zhang et al. [6, 7] have calculated the equilibrium distances and the hydrogen molecular ion $H_2^+$ in both ground and low-lying states in the presence of a magnetic field. The hydrogen molecular ion in a strong magnetic field has been studied for arbitrary orientations of the molecular axis in the non-aligned case by using the Lagrange-mesh method to obtain highly accurate results under these assumptions at various field strengths [8]. Zhang *et al.* [9] applied a simple one-center method based on B-spline basis sets in both radial and angular directions to calculate the equilibrium distances and the energies for the $H_2^+$ molecular ion in strong magnetic fields. Using the time-dependent density functional theory the variations in electron density and bonding have been investigated for the lowest $1\sigma_g$ state of the hydrogen molecule under strong magnetic fields [10]. Song et al. [11] have calculated the electronic structure and properties of the hydrogen molecule $H_2$ for the lowest $1\sigma_g$ and $1\sigma_u$ state in parallel magnetic fields using a full configuration-interaction (CI) method which is based on the Hylleraas-Gaussian basis set.

The aim of the present paper is to study the total energies, the dissociation energies and the binding energies of the hydrogen molecule $H_2$ and the hydrogen molecular ion $H_2^+$ in the presence of external magnetic field in framework of the Variational Monte Carlo (VMC) method, which is used widely in Refs [12-21]. Therefore, we have extended our previous work [22] to investigate the molecular systems using the VMC method.

## 2. The Trial Wave Functions

In this section, we introduce the trial wave functions that are used in our calculations. For the hydrogen molecular ion $H_2^+$ the following wave function is used:

$$\psi_1 = \sum_i C_i \, \lambda^{m_i} \, \mu^{n_i} \, exp(-\omega\lambda), \qquad (2.1)$$

where $\omega$ is a nonlinear parameter, $C_i$ are variational parameters and $m_i$ are positive or negative integers. Since the $1s\sigma_g$ ground-state has a gerade symmetry $n_i$ should be zero or a positive even integer. This type of wave function is proposed by Ishikawa *et al.* [23] to calculate the ground state of $H_2^+$ at different orders and the first excited state $1s\sigma_u$ in the absence of magnetic field using the iterative complement interaction (ICI) method. They have obtained accurate results compared to the corresponding exact values. The present calculations for the hydrogen molecule are based on the accurate trial wave function which take the form [24]:

$$\psi_2 = \sum_i C_i (1 + p_{12}) \, exp[-\alpha(\lambda_1 + \lambda_2)] \, \lambda_1^{m_i} \lambda_2^{n_i} \mu_1^{j_i} \mu_2^{k_i} \rho^{l_i}, \qquad (2.2)$$

where $p_{12}$ is an electron exchange operator and $C_i$ and $\alpha$ are the variational parameters which are calculated with the variational principle. The powers $m_i$ and $n_i$ of the variables $\lambda_1$ and $\lambda_2$ are positive or negative integers. Since the $1s\sigma_g$ ground-state has a gerade symmetry $n_i$ should be zero or a positive even integer. In the absence of magnetic field, $\psi_2$ gives a very accurate value for the ground state energy of hydrogen molecule [24].

## 3. Method of the Calculations

The present calculations are based on using variational Monte Carlo (VMC) method which is considered as one of the most important quantum Monte Carlo methods. It is based on a combination of two ideas namely the variational principle and the Monte Carlo evaluation of integrals using importance sampling based on the Metropolis algorithm [25].

The VMC methods are used to compute quantum expectation values of an operator with a given trial wave function, which is given in the previous section. The variational principle states that the expectation value of the Hamiltonian $H$ with respect to the trial wave function $\psi_T$ ($\psi_1$ or $\psi_2$) is the varitional energy [26]:

$$E_{VMC} = \frac{\int \psi_T^*(R)\, H\, \psi_T(R)\, dR}{\int \psi_T^*(R)\, \psi_T(R)\, dR} \geq E_{exact}, \qquad (3.1)$$

where $R$ is the $3N$-dimensional vector of the electron coordinates and $E_{exact}$ is the exact value of the energy of that state. Also, it is important to calculate the standard deviation of the energy [26]

$$\sigma = \sqrt{\frac{\langle E_L^2 \rangle - \langle E_L \rangle^2}{L\,(N-1)}} \qquad (3.2)$$

For more details concerning the VMC method, see Refs. [15,16]

## 4. The Hamiltonian of the System

In present work, we assume that the nuclear mass is infinite so that the calculations will be one in frame of the Born–Oppenheimer approximation and the magnetic field is oriented along the $z$ axis. The Schrödinger equation for the system can be written as follows

$$\widehat{H}\psi(r) = E\psi(r), \qquad (4.1)$$

where $r = (x, y, z)$ is the coordinate of each electron with respect to the center of mass of the two nuclei. The non-relativistic Hamiltonian $\widehat{H}$ for the hydrogen molecular system in a magnetic field, of strength B, can be written as [8, 27]:

$$H = -\frac{1}{2}\sum_{i=1}^{2} \nabla_i^2 - \sum_{i=1}^{2}\left(\frac{Z_a}{r_{ia}} + \frac{Z_b}{r_{ib}}\right) + \frac{Z_a Z_b}{r_{12}} + \frac{Z_a Z_b}{R} + H_M \qquad (4.2)$$

In the above equation $r_{ia(b)} = |\mathbf{r}_{ia(b)}|$ denotes the distance from electron '$i$' ($i = 1, 2$) to nucleus '$a$' ('$b$'), the charge parameters $Z_a = Z_b = 1$, $r_{12}$ is the interelectronic distance, $R$ is the internuclear distance and $H_M$ represents the magnetic part. The magnetic term takes the form

$$H_M = \begin{cases} \dfrac{\gamma^2}{8}\rho^2 & \text{for } H_2^+ \\ \dfrac{\gamma^2}{2}\rho^2 + \gamma(L_z + 2S_z) & \text{for } H_2 \end{cases} \qquad (4.3)$$

where $\gamma = \dfrac{B}{B_0}$ ($B_0 \approx 2.35 \times 10^5$ T) is the magnetic field strength, $L_z$ is the $z$-component of the total angular momentum, $S_z$ is the $z$-component of the total spin and $\rho^2 = \sum_{i=1}^{n}(x_i^2 + y_i^2)$. The index $n$ runs over the numbers of the electrons. Then, for hydrogen molecular ion $H_2^+$, $n = 1$ and the third term in (4.2) is absent, and in the case of hydrogen molecule, $H_2$, $n = 2$.

Equation (4.2) is best treated in the system of prolate spheroidal coordinates $(\lambda, \mu, \varphi)$ where $\varphi$ is the azimuthal angle and $\lambda$ and $\mu$ are defined by

$$\lambda = \frac{r_1 + r_2}{D}, \qquad \mu = \frac{r_1 - r_2}{D} \qquad (4.4)$$

In these coordinates, the kinetic-energy operator is written as

$$-\frac{1}{2}\nabla_i^2 = -\frac{2}{D^2\left(\lambda_i^2 - \mu_i^2\right)}\left\{\frac{\partial}{\partial \lambda_i}\left(\lambda_i^2 - 1\right)\frac{\partial}{\partial \lambda_i} + \frac{\partial}{\partial \mu_i}\left(1 - \mu_i^2\right)\frac{\partial}{\partial \mu_i}\right\} \qquad (4.5)$$

Figures 1 and 2 represent illustrations for the hydrogen molecular ion and the hydrogen molecule.

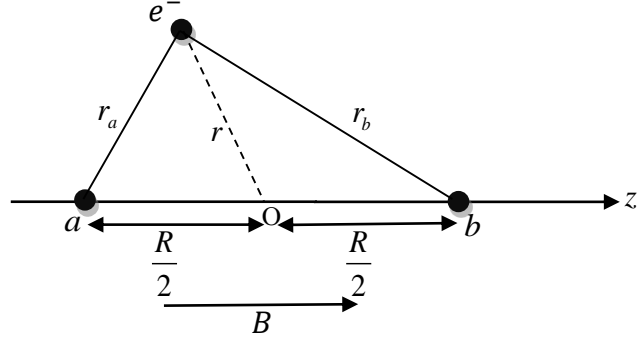

Figure-1 Geometrical illustration for the hydrogen molecular ion $H_2^+$ placed in a magnetic field directed along the $z$-axis. The protons are situated at a distance $R$ from each other.

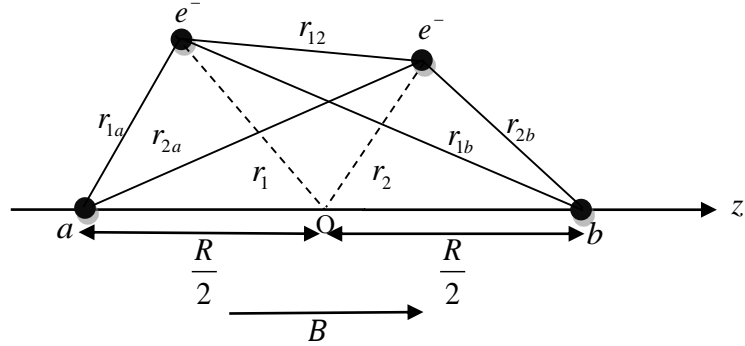

Figure-2 Geometrical illustration for the hydrogen molecule $H_2$ placed in a magnetic field directed along the $z$-axis. The protons are situated at a distance $R$ from each other.

## 5- Discussion of Results

The variational Monte Carlo method has been employed to calculate the ground state of the hydrogen molecular ion $H_2^+$ and the hydrogen molecule $H_2$ in the magnetic field regime between 0 a.u. and 10 a.u. All energies are obtained in atomic units i.e. ($\hbar = e = m_e = 1$) with set of $4 \times 10^7$ Monte Carlo integration points in order to make the statistical error as low as possible. The magnetic field is taken in parallel configuration, i.e. the angle between the molecular axis and the magnetic field direction is zero, $\theta = 0^o$ as shown in Figures 1 and 2. In the absence of magnetic field, our results for the total energy of the hydrogen molecular ion $H_2^+$ in the lowest state $(1s\sigma_g)$ equals -0.6023424 at the equilibrium distance of $R = 1.9972$ a.u, where for hydrogen molecule $H_2$ the total energy equals -1.173427 at $R = 1.40$ a.u.

In the present paper we have calculated the total energies, the binding energies and the dissociation energies of the $(1s\sigma_g)$ state as functions of the magnetic field over various field-strength regimes from (0 - 10 a.u.). The total energy $E_T$ of the hydrogen molecular ion $H_2^+$ and the hydrogen molecule $H_2$ are defined as the total electronic energy plus the repulsive energy between the nuclei: $E_T = E_{ele} + \frac{Z_a Z_b}{R}$. The least energy is required to produce a free electron and two nuclei, all infinitely far from each other in the presence of the field has been defined as the binding energy for the hydrogen molecular ion $H_2^+$, $E_b = \frac{\gamma}{2} - E_T$ by Larsen in [28]. For the hydrogen molecular ion $H_2^+$, the dissociation energy is defined as the least energy required to dissociate the molecule into one nucleus and one hydrogen atom in a magnetic field, $E_d = E_H - E_T$ where $E_H$ is the total energy of the hydrogen atom in a magnetic field, i.e. the product of the process $H_2^+ \rightarrow H + p^+$ [28].

In the case of the hydrogen molecule $H_2$ the product is $H_2 \rightarrow H(1s) + H(1s)$, which means that the energy in the dissociation limit corresponds to the energy of two hydrogen atoms in the lowest electronic state with positive $z$ parity, i.e. the quantity $E_d = E_T - \lim_{R \to \infty} E_T$. The binding energy is equal to the ionization energy and is always greater than the dissociation energy. The presence of free electron in the orbitals of molecules causes an induced magnetic field, to be produced and also due to the spin motion of this free electron. We observe this in the hydrogen molecular ion $H_2^+$ where, one electron presented in the $1s$ orbital will cause a weak induced magnetic field by the spin motion of this electron either clockwise or counterclockwise. Also, we show that there is no induced magnetic field in the hydrogen molecule $H_2$ as two electrons are paired in $1s$ orbital. The strength of the magnetic field is directly proportional to the odd number of free electrons presented in the orbitals.

The present calculations in the presence of a magnetic field are based on using the trial wave functions $\psi_1$ and $\psi_2$ are given by Eq. (2.1) and Eq. (2.2), respectively. For hydrogen molecular ion, the total energies are obtained by solving the Schrödinger equation given by Eq. (4.1) using $\psi_1$. Table-1 represents the obtained results for $H_2^+$. In our calculations the values of $E_H$ are taken from Ref [3]. In a similar way, the total energies and the dissociation energies are presented in Table-2 as functions of the magnetic field over various field-strength regimes for the ground $1s\sigma_g$ state of the hydrogen molecule $H_2$. From Tables-1 and -2, we observe that the binding energy, the dissociation energy and the total energy increase under an increase in the magnetic field strength. This is due to the fact that an increase in the field strength leads to increase in the movement of the electron and the electronic spatial distribution will be strongly confined in a smaller space. The probability of finding the electron in the region confined by the two nuclei becomes larger under increase of the field strength. So, the changes in the electronic properties of the ground state should be attributed to the increased electron density in the region between the nuclei centered at z = 0.

An interesting and general phenomenon for molecules is the decrease of the internuclear bond-length as the field strength increases. The decrease in the equilibrium internuclear distance originates from the simultaneous decrease of the electron clouds perpendicular and parallel to the magnetic field. This means that this state is the most tightly bound state for all magnetic field strengths because the electrons in this state are much closer to the nuclei than in the case of a free state. This increases the binding due to the attractive nuclear potential energy. Because the dissociation energy may be used to measure the stability of a molecular system in a magnetic field, it is useful to explore the behavior of this quantity under increasing field strength. There is an increase in the binding energy of molecular systems as the magnetic field strength gets larger. An increase in the binding energy comes from the result of strong localization of the electrons around the nuclei. The electron-electron Coulomb repulsion is taken into account.

It is interesting to compare our work with the previous works using different wave functions and different methods such as [3, 4, 7, 30, 31] for the ground state $1s\sigma_g$ of the $H_2^+$ molecular ion and [11, 35, 36] for the ground state $1s\sigma_g$ of the hydrogen molecule $H_2$. It is clear that our results are in good agreement with the pervious data.

Figures-3 and -4 show the variation of the ground state energy of the $H_2^+$ molecular ion and the hydrogen molecule $H_2$ in the presence of a magnetic field from $\gamma = 0.0$ to $\gamma = 10.0$, respectively, versus the internuclear distance $R$. These show that when the magnetic field strength increases the ground state energy increases as seen from the change in the values of the ground state energy from $\gamma = 0.0$ to $\gamma = 10.0$. We can observe that at $\gamma = 10.0$ in the both cases the value of the total energy is the highest energy. At $\gamma = 0.0$ the total energy of the hydrogen molecular ion $H_2^+$ is larger than the hydrogen molecule because of the induced magnetic field.

**Table (1):** Total energy $E_T$, binding energy $E_b$ and dissociation energy $E_d$ of the ground state $1s\sigma_g$

| $\gamma$ | $R_{eq}$ | References | $E_T$ | $E_b$ | $E_d$ |
|---|---|---|---|---|---|
| 0.0 | 1.9971934 | This work [4] | -0.602501700(4) −0.6026346191066 | 0.602501700 - | 0.102501700 - |
| | 1.997193 | This work [3] | -0.602501700(4) −0.602634619 | 0.602501700 0.602634619 | 0.102501700 0.102634619 |
| | 1.9971 | This work [30] | -0.602390800(2) −0.602625 | 0.602390800 0.602625 | 0.102390800 - |
| 0.002 | 1.99719 | This work [31] | -0.602508600(1) −0.60263398 | 0.603508600 0.60363398 | 0.102509600 0.10263498 |
| 0.008 | 1.997162 | This work [3] | -0.602477700(2) −0.602624361 | 0.606477700 0.606624361 | 0.102493699 0.102640360 |
| | 1.99716 | This work [31] | -0.602610500(1) −0.60262436 | 0.606610500 0.60662436 | 0.102626499 0.10264036 |
| 0.02 | 1.996991 | This work [3] | -0.602479700(2) −0.602570515 | 0.612479700 0.612570515 | 0.102579656 0.102670471 |
| 0.1 | 1.992212 | This work [3] | -0.600764400(0) −0.601038207 | 0.6507644 0.651038207 | 0.10323792 0.103511727 |
| | 1.9922107 | This work [4] | -0.600734200(0) −0.6010382074075 | 0.6507342 - | 0.10320772 - |
| | 1.99221 | This work [31] | -0.600685200(0) −0.60103820 | 0.6506852 0.65103820 | 0.10315872 0.10351172 |
| 0.425434 | 1.924 | This work [29] | -0.575015700(2) −0.575359 | 0.7877327 0.788125 | 0.114587838 - |
| | 1.9234 | This work [3] | -0.574840800(1) −0.575370830 | 0.787565 0.788087830 | 0.114412938 0.114942968 |
| | 1.9 | This work [32] | -0.574771100(3) −0.575075 | 0.7874881 0.763412 | 0.114343238 0.114685 |

of the $H_2^+$ molecular ion in a parallel magnetic field from $\gamma = 0.0$ to $\gamma = 10.0$. Note that the present definition for $E_T$ is equivalent to the definition of $E_e$ in Wille's work [29]. In parentheses, we show the statistical error in the last figure.

Table-1 Continued

| $\gamma$ | $R_{eq}$ | References | $E_T$ | $E_b$ | $E_d$ |
|---|---|---|---|---|---|
| 1.0 | 2.0 | This work<br>[4]<br>[7] | -0.470237300(5)<br>−0.47054001262014<br>−0.4705400126203 | 0.9702373<br>-<br>- | 0.139068403<br>-<br>- |
|  | 1.752084 | This work<br>[7]<br>[4] | -0.468984500(5)<br>−0.474988245275<br>−0.474988245275 | 0.9689845<br>-<br>- | 0.137815603<br>-<br>- |
|  | 1.7520838 | This work<br>[4] | -0.468984500(5)<br>−0.474988245275 | 0.9689845<br>- | 0.137815603<br>- |
|  | 1.7521 | This work<br>[3] | -0.468523800(8)<br>−0.474988245 | 0.9685238<br>0.974988245 | 0.1373549033<br>0.143819348 |
|  | 1.752 | This work<br>[28] | -0.468317700(7)<br>−0.4749 | 0.9683177<br>0.9749 | 0.137148803<br>- |
| 2.12717 | 1.5025 | This work<br>[3] | -0.17393060(3)<br>−0.174910873 | 1.2375156<br>1.238495873 | 0.1952071163<br>0.153634357 |
|  | 1.448 | This work<br>[33] | -0.1520147(1)<br>−0.15225 | 1.2155997<br>1.215835 | 0.1732912163<br>0.2570 |
| 3.0 | 1.376 | This work<br>[30] | 0.10122530(7)<br>0.10485 | 1.3987747<br>1.39515 | 0.2342417107<br>- |
|  | 1.3754 | This work<br>[3] | 0.10123410(7)<br>0.104455347 | 1.3987659<br>1.39554465 | 0.2342329107<br>0.231011664 |
| 4.25434 | 1.2465 | This work<br>[3] | 0.5304074(1)<br>0.544794264 | 1.5967626<br>1.582375737 | 0.2892954573<br>0.274908594 |
|  | 1.2464 | This work<br>[30] | 0.5305169(1)<br>0.544895 | 1.5966531<br>1.5823 | 0.2891859573<br>- |
|  | 1.246 | This work<br>[29] | 0.530954(1)<br>0.545154 | 1.596216<br>1.582016 | 0.2887488573<br>- |
| 10.0 | 0.957 | This work<br>[3] | 2.8221030(5)<br>2.825014 | 2.177897<br>2.174986 | 0.430107<br>0.427196 |
|  | 0.950 | This work<br>[34]<br>[29] | 2.827940(4)<br>2.8327<br>2.8250 | 2.17206<br>2.1673<br>2.1750 | 0.42427<br>0.41955<br>- |

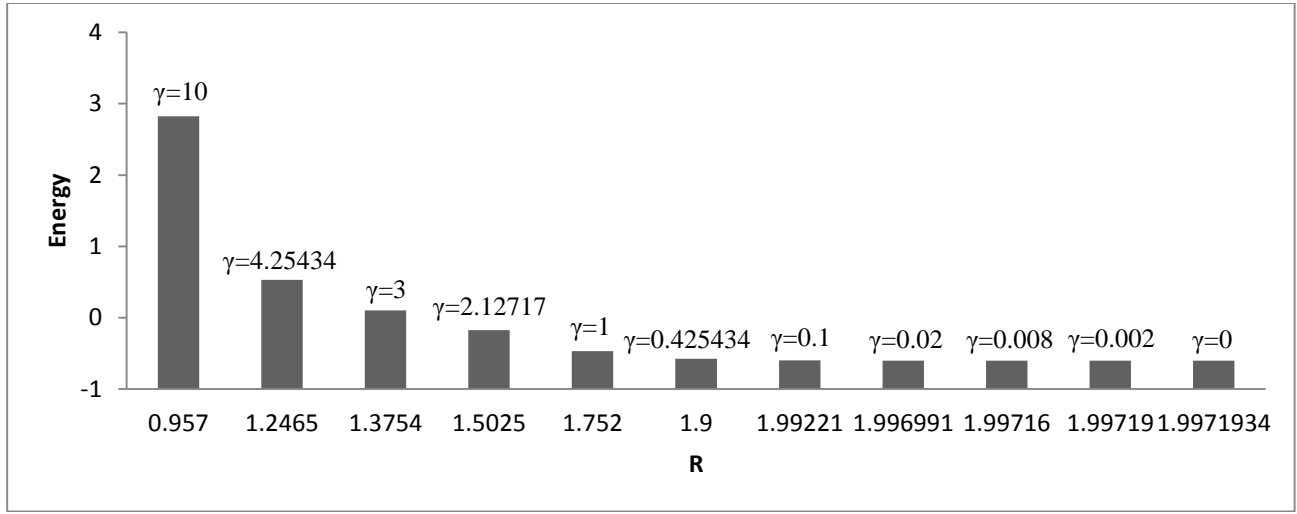

**Fig. 3**: Ground-state energy of the hydrogen molecular ion $H_2^+$ in the presence of a magnetic field from $\gamma = 0.0$ to $\gamma = 10.0$ versus the internuclear distance $R$.

**Table-2**: Total energy $E_T$, and dissociation energy $E_d$ of the ground state $1s\sigma_g$ of the hydrogen molecule $H_2$ in a parallel magnetic field from $\gamma = 0.0$ to $\gamma = 10.0$. In parentheses, we show the statistical error in the last figure.

| $\gamma$ | $R_{eq}$ | References | $E_T$ | $E_d$ |
|---|---|---|---|---|
| 0.0 | 1.40 | This work | -1.173427(1) | 0.173429 |
| | | [35] | −1.173436 | 0.173438 |
| | | [11] | −1.1744477 | 0.1744478 |
| 0.001 | 1.40 | This work | -1.17342(1) | 0.173422 |
| | | [35] | −1.173436 | 0.173438 |
| 0.005 | 1.40 | This work | -1.173285(1) | 0.173301 |
| | | [35] | −1.173424 | 0.173440 |
| 0.01 | 1.40 | This work | -1.172867(1) | 0.1729171 |
| | | [35] | −1.173396 | 0.173450 |
| | | [11] | −1.1744096 | 0.1744597 |
| 0.05 | 1.40 | This work | -1.171858(1) | 0.1731047 |
| | | [35] | −1.172407 | 0.173658 |
| | | [11] | −1.173497 | 0.1747437 |
| 0.1 | 1.40 | This work | -1.170512(7) | 0.1754591 |
| | | [11] | −1.1706617 | 0.1756088 |
| | 1.39 | This work | -1.166586(5) | 0.171542 |
| | | [35] | −1.169652 | 0.174608 |
| 0.2 | 1.39 | This work | -1.135232(1) | 0.154469 |
| | | [35] | −1.158766 | 0.178001 |
| | | [11] | −1.159579 | 0.178816 |
| 0.4254414 | 1.349 | This work | -1.110247(2) | - |
| | | [35] | −1.110362 | |
| | 1.337 | This work | -1.079712(1) | - |
| | | [36] | −1.0822 | |
| 0.5 | 1.33 | This work | -1.078379(2) | 0.183958 |
| | | [35] | −1.089082 | 0.194663 |
| | | [11] | −1.089750 | 0.195329 |
| 1.0 | 1.23 | This work | -0.8866(7) | 0.224262 |
| | | [11] | −0.891184 | 0.228846 |
| 2.0 | 1.09 | This work | -0.3334(2) | 0.288972 |
| | | [35] | −0.335574 | 0.291170 |
| | | [11] | −0.336236 | 0.291808 |
| 2.127207 | 1.07 | This work | -0.2501(1) | - |
| | | [35] | −0.255591 | |
| 4.254414 | 0.898 | This work | 1.2322(5) | - |
| | | [35] | 1.233808 | |
| | 0.859 | This work | 1.3266(4) | - |
| | | [36] | 1.3326 | |
| 5.0 | 0.86 | This work | 1.8077(5) | 0.4459257 |
| | | [35] | 1.801212 | 0.438015 |
| | | [11] | 1.8004883 | 0.438714 |
| 10.0 | 0.70 | This work | 5.8826(3) | 0.6105216 |
| | | [35] | 5.889023 | 0.615473 |
| | | [11] | 5.8882422 | 0.6161638 |

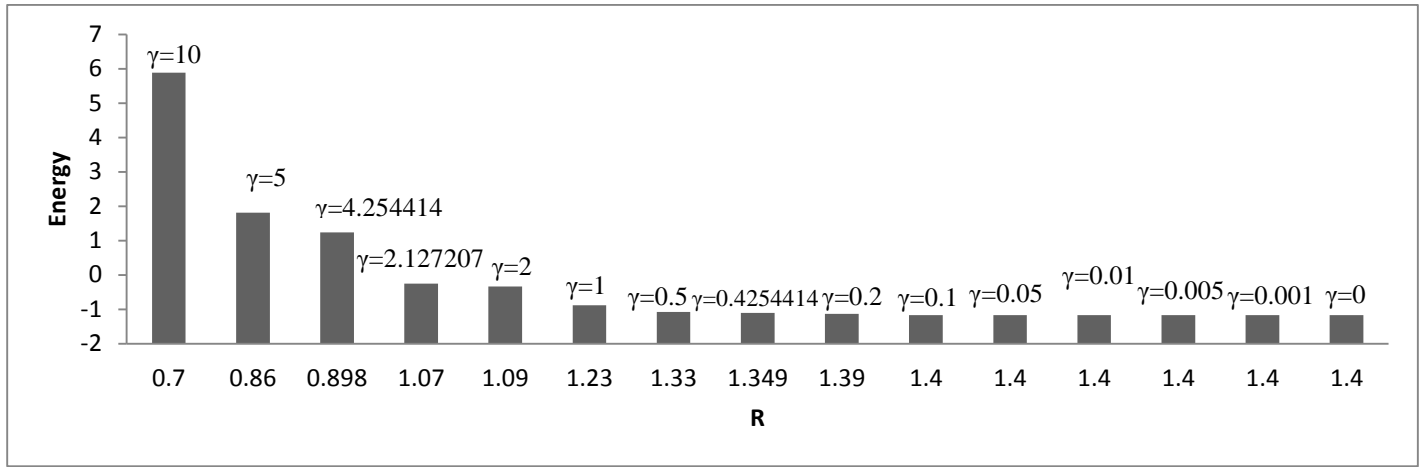

Fig. 4 Ground-state energy of the hydrogen molecule $H_2$ in the presence of a magnetic field from $\gamma = 0.0$ to $\gamma = 10.0$ versus the internuclear distance $R$.

## 6. Conclusions

In the present work, we have studied the hydrogen molecular ion $H_2^+$ and the hydrogen molecule $H_2$ in the presence of a magnetic field by using the well-known Variational Monte Carlo method. In present study, the molecular axis is usually aligned along the field axis. Accordingly, we have calculated the total energies and the dissociation energies with respect to the magnetic field for both of the hydrogen molecular ion $H_2^+$ and the hydrogen molecule $H_2$ and the binding energies for the hydrogen molecular ion $H_2^+$ only by using two accurate trial wave functions of the $(1s\sigma_g)$ state over various field-strength regimes. While increasing the field strength, the equilibrium distance $R_{eq}$ decreases and the total energy, the dissociation energy and binding energy $E_b$ increase monotonously. In both cases our results exhibit a good accuracy under various field strengths comparing with previous values obtained by using different methods and different forms of trial wave functions. This is due to the fact that we have used two trial wave functions each of them takes into consideration the electron-electron correlation. Finally, we conclude that the applications of VMC method can be extended successfully to cover the case of molecules under the effect of the magnetic field.